\documentstyle[epsf,multicol,aps]{revtex}
\begin{document}
\draft

\title{Competition of electron capture and beta-decay rates in
  supernova collapse}

\author{G. Mart\'{\i}nez-Pinedo$^1$, K. Langanke$^1$ and
  D.J. Dean$^2$}

\address{$^1$Institute for Physics and Astronomy, University of
  Aarhus, Denmark and
  Theoretical Astrophysics Center, University of Aarhus, Denmark\\
  $^2$Physics Division, Oak Ridge National Laboratory, Oak Ridge,
  Tennessee 37831-6373 and Department of Physics and Astronomy,
  University of Tennessee, Knoxville, Tennessee 37996 USA}

\date{\today} 

\maketitle

\begin{abstract}
  We calculate supernova electron capture and $\beta$ decay rates for
  various $pf$-shell nuclei using large-scale shell model techniques.
  We show that the centroid of the Gamow-Teller strength distribution
  has been systematically misplaced in previous rate estimates. Our
  total electron capture rates are significantly smaller than
  currently adopted in core collapse calculations, while the total
  $\beta$ decay rates change less. Our calculation shows that for
  electron-to-baryon ratios $Y_e=0.42$-0.46 $\beta$ decay rates are
  larger than the competing electron capture rates.
\end{abstract}


\begin{multicols}{2}

\section{Introduction}

Weak interaction processes play a decisive role in the early stage of
the core collapse of a massive star \cite{Bethe78,Bethe90}. First,
electron capture on nuclei in the iron mass region, starting after the
core mass exceeds the appropriate Chandrasekhar mass limit, reduces
the electron pressure, thus accelerating the collapse, and lowers the
electron-to-baryon ratio, $Y_e$, thus shifting the distribution of
nuclei present in the core to more neutron-rich material. Second, many
of the nuclei present can also $\beta$ decay. While this process is
quite unimportant compared to electron capture for initial $Y_e$
values around 0.5, it becomes increasingly competative for
neutron-rich nuclei due to an increase in phase space related to
larger $Q_\beta$ values. However, $\beta$ decay on nuclei with masses
$A>60$ have not yet been considered in core collapse studies
\cite{Thielemann}.  This is surprising since Gerry Brown pointed out
nearly a decade ago \cite{Brown89} that certain nuclei heavier than
$A=60$, like $^{63}$Co and $^{64}$Co, have very strong $\beta$-decay
matrix elements making it conceivable that they can actually compete
with electron capture.  Brown argued that this might have quite
interesting consequences for the collapse. During this early stage of
the collapse, neutrinos produced in both electron capture and $\beta$
decay still leave the star. Therefore, a strong $\beta$-decay rate
will cool the star without lowering the $Y_e$ value. As a consequence,
the $Y_e$ value at the formation of the homologous core (after
neutrino trapping) might be larger than assumed. This results in a
smaller envelope, and less energy is required for the shock to travel
through the material.

Following Brown's suggestion, $\beta$ decay rates for nuclei in the
mass range $A=48-70$ were investigated \cite{Aufderheide}.  These
studies were based on the same strategy and formalism as already
employed by the pioneering work in this field by Fuller, Fowler and
Newman (commonly abbreviated by FFN) \cite{FFN}. The important idea in
FFN was to recognize the role played by the Gamow-Teller resonance in
$\beta$ decay. Other than in the laboratory, $\beta$ decay rates under
stellar conditions are significantly increased due to thermal
population of the Gamow-Teller back resonance in the parent nucleus
(the GT back resonance are the states reached by the strong GT
transitions in the inverse process (electron capture) built on the
ground and excited states, see \cite{FFN,Aufderheide}) allowing for a
transition with a large nuclear matrix element and increased phase
space. Indeed, Fuller et al. concluded that the $\beta$ decay rates
under collapse conditions is dominated by the decay of the back
resonance. In a more recent work, Aufderheide et al. came to the same
conclusion. Inspired by the independent particle model, the authors of
Ref. \cite{Aufderheide} estimated the $\beta$ decay rates in a similar
fashion as the electron capture rates, and phenomenologically
parametrized the position and the strength of the back resonance. This
estimate was supplemented by an empirical contribution, placed at zero
excitation energy, which simulates low-lying transition strength
missed by the GT resonance.

When extending the FFN rates to nuclei with $A>60$, Aufderheide et al.
found indeed that the $\beta$ decay rates are strong enough to balance
the electron capture rates for $Y_e \approx 0.42-0.46$. Nevertheless,
these results have never been explored in details in core collapse
calculations.

In recent years the parametrization of the electron capture and
$\beta$ decay rates as adopted by FFN and Aufderheide et al. have
become questionable due to experimental data
\cite{gtdata1,gtdata2,gtdata3,gtdata4,gtdata5} and have been
critizised on the basis of more elaborate theoretical models
\cite{Aufderheide96,Dean98,Langanke98a,Langanke98b}. We will show in
this paper, that although the previous weak interaction rates for core
collapse are systematically incorrect, the important observation that
electron and $\beta$ decay rates balance each other for a certain
range of $Y_e$ values is indeed correct. Our conclusions will be based
on large-scale shell model calculations for several key nuclei which,
due to Aufderheide et al. \cite{Aufderheide}, contribute most
significantly to the electron capture and $\beta$ decay rates at
various stages of the collapse. These shell model calculations
reproduce the measured GT strength distributions for nuclei in the
mass range $A=50-64$ very well \cite{Martinez99}.  Furthermore modern
large-scale shell model calculations also agree with measured
half-lifes very well. Thus for the first time one has a tool in hand
which allows for a reliable calculation of presupernova electron
capture and $\beta$ decay rates. Modern shell model calculations come
in two varieties: large-scale diagonalization approaches
\cite{Caurier98} and shell model Monte Carlo (SMMC) techniques
\cite{Johnson,Koonin}.  The latter can treat the even larger model
spaces, but has limitations in its applicability to odd-A and odd-odd
nuclei at low temperatures, which does not apply to the former.  More
importantly the diagonalization approach allows for detailed
spectroscopy, while the SMMC model yields only an ``averaged'' GT
strength distribution which introduces some inaccuracy into the
calculation of the capture and decay rates.

We will consistenly use in the following the shell model
diagonalization approach to study these rates.  Due to the very large
m-scheme dimensions involved, the GT strength distributions have been
calculated in truncated model spaces which fulfill the Ikeda sum rule.
However, at the chosen level of truncation involving typically 10
million configurations or more, the GT strength distribution is
virtually converged.  As residual interaction we adopted the recently
modified version of the KB3 interaction which corrects the slight
inefficiencies in the KB3 interaction around the $N=28$ subshell
closure \cite{Ni56}.  In fact the modified KB3 interaction i)
reproduces all measured GT strength distributions very well and ii)
describes the experimental level spectrum of the nuclei studied here
quite accurately \cite{Martinez99,Ni56}.  As $0\hbar\omega$ shell
model calculations, i.e.  calculations performed in one major shell,
overestimate the experimental GT strength by a universal factor
\cite{Wildenthal,Langanke95,Martinez96}, we have scaled our GT
strength distribution by this factor, $(0.74)^2$.

Large-scale shell model calculations of the electron capture rates for
key nuclei in the presupernova collapse have been reported already
elsewhere \cite{Langanke98a,Langanke98b}; see also the SMMC results in
\cite{Dean98}.  In these studies it became apparent that the
phenomenological parametrization of the GT contribution to the
electron capture and $\beta$ decay rates, as introduced in FFN
\cite{FFN} and subsequently used by Aufderheide et al.
\cite{Aufderheide}, is systematically incorrect.  These authors have
placed the centroid of the GT strength distribution at too high an
excitation energy in the daughter nucleus for electron capture on
even-even nuclei, while for capture on odd-A and odd-odd nuclei they
underestimated the energy of the GT centroid noticeably.  For capture
on even-even nuclei this has comparably little effect as FFN
overcompensate the misplacement of the GT centroid by a too large
empirical contribution at zero excitation energy. Typically the FFN
rates are roughly a factor of 5 larger than the shell model rates for
capture on even-even nuclei like $^{56,58}$Ni or $^{58}$Fe. For
capture on odd-odd and odd-A nuclei the misplacement of the GT
centroid makes the FFN rates about 1-2 orders of magnitude too large
compared to the shell model rates. As a consequence, FFN and
Aufderheide et al. have noticeably overestimated the electron capture
rates for the early stage of the supernova collapse.

Which consequences do the misplacement of the GT centroids have for
the competing $\beta$ decays? In odd-A and even-even nuclei (the
daughters of electron capture on odd-odd nuclei), experimental data
and shell model studies place the back-resonance at higher excitation
energies than assumed by FFN and Aufderheide et al.
\cite{Aufderheide}. Correspondingly, its population becomes less
likely at the temperatures available during the early stage of the
collapse ($T_9 \approx 5$, where $T_9$ measures the temperature in
$10^9$ K) and hence the contribution of the back-resonance to the
$\beta$ decay rates for even-even and odd-A nuclei decreases.  Due to
Aufderheide et al. \cite{Aufderheide}, some of the most important
$\beta$ decay nuclei (defined by the product of abundance and $\beta$
decay rate and listed in Tables 18-22 in \cite{Aufderheide}) are
odd-odd nuclei. For these nuclei, all available data, stemming from
(n,p) reaction cross section measurements on even-even nuclei like
$^{54,56,58}$Fe or $^{58,60,62,64}$Ni, and all shell model
calculations indicate that the back-resonance resides actually at
lower excitation energies than previously parametrized. Consequently,
the contribution of the back-resonance to the $\beta$ decay rate of
odd-odd parent nuclei should be larger than assumed in the
compilations. We note that this general expectation has already been
conjectured in Ref.  \cite{Aufderheide96} on the basis of (n,p) data
available at that time.  These authors have attempted to fit the data
within a strongly truncated shell model calculation which then in turn
has been used to predict a corresponding $\beta$ decay rate. This
procedure is viewed as rather uncertain as i) the large energy
resolution in the data made its convolution into a $\beta$ decay rate
imprecise and ii) the shell model truncation level was too inaccurate
in order to estimate reliably the contribution of other states than
the back-resonance to the decay rate.  These shortcomings can be
overcome in recent state-of-the-art large scale shell model
calculations. We have calculated the $\beta$ decay rates for several
nuclei under relevant core collapse conditions ($\rho_7=10-1000$,
where $\rho_7$ measures the density in $10^7$ g/cm$^3$ and
temperatures $T_9=1-10$).  These nuclei include even-even ones
($^{52}$Ti,$^{54}$Cr, $^{56,58,60}$Fe), odd-A nuclei ($^{59}$Mn,
$^{59,61}$Fe, $^{61,63}$Co) and odd-odd nuclei
($^{50}$Sc,$^{54,56}$Mn, $^{58,60}$Co).  The selection has been made
to include those nuclei which have been ranked as most important for
core collapse simulations by Aufderheide et al. \cite{Aufderheide}. In
fact, with the rates of Ref. \cite{Aufderheide} these 15 nuclei
contribute between $65 \%$ and $86 \%$ to the change of $Y_e$ due to
$\beta$ decay in the range $Y_e=0.44-0.47$.

Although the formula for the presupernova $\beta$ decay rate
$\lambda_{\rm \beta}$ is well known (e.g. \cite{FFN,Aufderheide}), we
have chosen to quote the basic result here as this allows for the
easiest discussion of the improvement incorporated in our calculation
compared to previous work. Thus,
\end{multicols}

\begin{equation}
\lambda_{\rm{\beta}}=\frac{\ln 2}{6163 \rm{sec}}
\sum_{ij} \frac{(2J_i+1) \exp{[-E_i/kT]}}{G}
S^{ij}_{\rm{GT}} 
\frac{c^3}{(m_e c^2)^5}\int_0^{\cal L} dp p^2 (Q_{ij}-E_e)^2
\frac{F(Z+1,E_e)}{1+\exp\left[kT(\mu_e-E_e)\right]}\;,
\end{equation}

\begin{multicols}{2}

\noindent
where $E_e$, $p$, and $\mu_e$ are the electron energy, momentum, and
chemical potential, and ${\cal L}=(Q_{if}^2 - m_e^2 c^4)^{1/2}$;
$Q_{if}=E_i-E_f$ is the nuclear energy difference between the initial
and final states, while $S^{ij}_{\rm {GT}}$ is their GT transition
strength.  $Z$ is the charge number of the parent nucleus, $G$ is the
partition function, $G=\sum_i (2J_i+1) \exp{[-E_i/kT]}$, while $F$ is
the Fermi function which accounts for the distortion of the electron's
wave function due to the Coulomb field of the nucleus.  The values for
the chemical potential are taken from \cite{Dean98}.

To estimate the rates at finite temperatures, the compilations
employed the so-called Brink hypothesis
\cite{Aufderheide,Aufderheide91} which assumes that the GT strength
distribution on excited states is the same as for the ground state,
only shifted by the excitation energy of the state.  We have not used
this approximation, but have performed shell model calculations for
the individual transitions. Our sum over initial states includes i)
explicitly the ground state and several excited states in the parent
nucleus (usually at least all levels below 1 MeV excitation energy)
and ii) all back-resonances which can be reached from the levels in
the daughter nucleus below 1 MeV excitation energy. As these
back-resonances also include parent states below 1 MeV, special care
has been taken in avoiding double-counting. The partition function is
consistenly summed over the same initial states.

Here a word of caution is in order. We have calculated the GT strength
distributions using 33 Lanczos iterations in all allowed angular
momentum and isospin channels. This is usually sufficient to converge
in the states at excitation energies below $E= 3$ MeV.  At higher
excitation energies, $E>3$ MeV, the calculated GT strengths represent
centroids of strengths, which in reality are split over many states.
While this does not introduce uncertainties in the summing over the GT
strengths (the numerator in (1)), it might be inconsistent for the
calculation of the partition function. However, this is practically
not the case, as at the rather low temperatures of concern here the
partition function is given by those states which have already
converged in our model space. Nevertheless there might be states
outside of our model space (intruder states) which will be missed in
our evaluation of the $\beta$ decay rates. But their statistical
weight in both numerator and denominator in the rate equation (1) is
small.  Although our calculations agree well with the experimental
informations available (excitation energies and GT transition
strengths), we have replaced the shell model results by data whenever
available.

In Fig. 1 we compare our shell model $\beta$ decay rates with those of
FFN for selected nuclei representing the three distinct classes:
even-even ($^{54}$Cr, $^{56,60}$Fe), odd-A ($^{59}$Mn, $^{57,59}$Fe)
and odd-odd ($^{54,56}$Mn, $^{58}$Co).  We note again that $\beta$
decay of the nuclei studied here is important at temperatures $T_9
\leq 5$ \cite{Aufderheide}.  For the odd-odd nuclei we calculate rates
similar to those of FFN. This approximate agreement is, however,
somewhat fortunate. In FFN the misplacement of the GT back-resonances
has been compensated by too large values for the total GT strengths
(FFN adopted the unquenched single particle estimate) and the
low-lying strengths.  At higher temperatures ($T_9>5$), the FFN rates
for odd-odd nuclei are larger than our shell model rates.  For odd-A
and even-even nuclei our shell model rates are significantly smaller
than the FFN rates as the back resonance, for $T_9 < 5$, is less
populated thermally than in the FFN parametrization. Using the FFN
rates, even-even nuclei were found to be unimportant for $\beta$ decay
in the core collapse; our lower rates make them even less important.
This situation is somewhat different for odd-A nuclei which (
$^{57,59}$Fe, $^{59}$Mn) have been identified as important in Ref.
\cite{Aufderheide} adopting the FFN rate. Aufderheide et al. have
added several odd-A nuclei with masses $A>60$ (which are not
calculated in FFN) to the list of those nuclei which significantly
change the $Y_e$ value during the collapse by $\beta$ decays. These
nuclei include $^{61}$Fe and $^{61,63}$Co; we will show below that
their rates have also been overestimated significantly in Ref.
\cite{Aufderheide}.  Our shell model rates indicate that the
importance of odd-A nuclei is significantly overestimated when the
previously compiled values are adopted.

In Ref. \cite{Aufderheide96} $\beta$ decay rates for several nuclei
have been estimated in strongly truncated shell model calculations, in
which these authors allowed a maximum of 1 nucleon to be excited from
the $f_{7/2}$ shell to the rest of the pf-shell in the daughter
nucleus, and fitted the single particle energy spectra to reproduce
measured (n,p) data for $^{54,56}$Fe, $^{58}$Ni and $^{59}$Co; the
(n,p) data constrain the back-resonance transition to the ground
states in the $\beta$ decays of $^{54,56}$Mn, $^{58}$Co and $^{59}$Fe.
Our shell model rates are compared to the estimates of Ref.
\cite{Aufderheide96} in Fig. 2. For the 3 odd-odd nuclei, the
agreement is usually better than a factor 2. This is due to the fact
that these rates are dominated by the back-resonances, i.e. the (n,p)
data of the daughter nucleus, which are reproduced in our large-scale
shell model approach and have been fitted in Ref.
\cite{Aufderheide96}.  For the odd-A nucleus $^{59}$Fe our $\beta$
decay rate is about an order of magnitude lower than the estimate of
Ref. \cite{Aufderheide96} at $T_9=2$, while the rates agree for $T_9
>6$, where it is dominated by the back-resonances. At the lower
temperatures, $\beta$ decays of low-lying states are important which
might be overestimated in the truncated calculation.

What might the revised $\beta$ decay rates mean for the core collapse?
To investigate this question we study the change of the
electron-to-baryon ratio, ${\dot Y}_e$, along a stellar trajectory.
Following Ref. \cite{Aufderheide}, we define
\begin{equation}
Y_e=\sum_k \frac{Z_k}{A_k} X_k
\end{equation}
where the sum runs over all nuclear species present in the core. $Z$,
$A$, and $X$ are the charge, mass number, and mass fraction of the
nucleus, respectively. The mass fraction is given by nuclear
statistical equilibrium \cite{Aufderheide}; we will use the values as
given in Tables 14-24 of Ref. \cite{Aufderheide}. Noting that $\beta$
decay ($\beta$) increases the charge by one unit, while electron
capture (ec) reduces it by one unit, we have
\begin{equation}
{\dot Y}_e^{ec(\beta)} = \frac{dY_e^{ec(\beta)}}{dt}=-(+) \sum_k
\frac{X_k}{A_k} \lambda_k^{ec(\beta)}
\end{equation}
where $\lambda_k^{ec}$ and $\lambda_k^{\beta}$ are the electron
capture and $\beta$ decay rates of nucleus $k$. For several key nuclei
we have calculated these rates within large-scale shell model studies.
Some of the results are listed in Tables 1 and 2, where they are also
compared to the FFN rates and the ones of Ref. \cite{Aufderheide}.
This comparison also includes the $\beta$ decay rates for $^{61}$Fe
and $^{61,63}$Co, which, due to \cite{Aufderheide} and earlier
suggested by Brown \cite{Brown89,Aufderheide90}, are important when
the stellar trajectory reaches electron-to-baryon values $Y_e
=0.44-0.46$. Our shell model rates agree for $^{63}$Co with the rate
of Aufderheide et al., but are smaller than the estimates of these
authors by factors 2 and 5 for $^{61}$Fe and $^{61}$Co, respectively.
We note that the strong ground state decay of $^{63}$Co contributes
about $15\%$ to the total rate at the condition listed in Table 2.
Some of the electron capture rates are taken from
\cite{Langanke98a,Langanke98b}, while several other shell model rates
are presented here for the first time (e.g. for $^{54,56}$Fe,
$^{58}$Ni and the odd-A nuclei).  Although the nuclei, for which
reliable shell model rates are now available, include the dominant
ones at the various stages of the early collapse (due to the ratings
in Ref. \cite{Aufderheide}), there are upto 250 nuclei present in NSE
at higher densities \cite{Aufderheide}. Although we are currently
working at a revised compilation for $\beta$ decay and electron
capture rates for nuclei in the mass range $A=45-65$, its completion
is computer-intensive and tedious. Nevertheless some important
conclusions can already be drawn from the currently available data.

At first we will follow the stellar trajectory as given in Ref.
\cite{Aufderheide}, although some comments about this choice are given
below. We estimated ${\dot Y}_e^{ec}$ and ${\dot Y}_e^{\beta}$
separately on the basis of the 25 most important nuclei listed in
Tables 14-24 in \cite{Aufderheide}. We used shell model rates for the
nuclei listed in Table 1 and 2.  For the other nuclei we scaled the
FFN rates using the following scheme which corrects for the systematic
misplacement of the GT centroid and is derived by the comparison of
FFN and shell model rates for the nuclei listed in Tables 1 and 2.
The FFN electron capture rates have been multiplied by 0.2
(even-even), 0.1 (odd-A) and 0.04 (odd-odd), while the FFN $\beta$
decay rates have been scaled by 0.05 (even-even), 0.025 (odd-A) and
1.5 (odd-odd). The results for ${\dot Y}_e^{ec,\beta}$ are plotted in
Fig. 3, where they are also compared to the values obtained for the
FFN rates. One observes that the shell model rates reduce ${\dot
  Y}_e^{ec}$ significantly, by more than an order of magnitude for
$Y_e<0.47$.  This is due to the fact, that, except for $^{56}$Ni, all
shell model electron capture rates are smaller than the
recommendations given in the FFN and Aufderheide et al. compilations
\cite{FFN,Aufderheide}.  In particular, this is drastic for capture on
odd-odd nuclei, which due to these compilations, dominate ${\dot
  Y}_e^{ec}$ at densities $\rho_7>10$. The shell model $\beta$ decay
rates also reduce ${\dot Y}_e^{\beta}$, however, by a smaller amount
than for electron capture. This is mainly caused by the fact that the
shell model $\beta$ decay rates of odd-odd nuclei are about the same
as the FFN rates or even slightly larger, for reasons discussed above.

It is interesting to note that FFN typically give higher $\beta$-decay
rates for odd-A nuclei than Aufderheide et al. \cite{Aufderheide},
while it is vice versa for odd-odd nuclei. As a consequence ${\dot
  Y_e^{\beta}}$ is dominated by odd-A nuclei for $Y_e<0.46$ if the FFN
rates are used, while odd-odd nuclei contribute significantly if the
rates of \cite{Aufderheide} are adopted.  In either case, both
compilations yield rather similar profiles for ${\dot Y}_e^{ec,\beta}$
(see Fig. 14 in \cite{Aufderheide}).  The important feature in Fig. 3
is the fact that the $\beta$ decay rates are larger than the electron
capture rates for $Y_e=0.42-0.455$, which is also already true for the
FFN rates \cite{Aufderheide}.

So far we have used the same stellar trajectory as in Ref.
\cite{Aufderheide}. This allowed a comparison with the conclusions
obtained in that reference. However, this assumption is inconsistent,
and, in fact, was already inconsistent in \cite{Aufderheide}. The
chosen stellar trajectory is based on runs performed with the stellar
evolution code KEPLER \cite{Kepler} which uses the FFN electron
capture rates, but quite outdated $\beta$ decay rates
\cite{Aufderheide94}, following the old belief that $\beta$ decay
rates are unimportant \cite{Aufderheide94}.  The outdated $\beta$
decay rates were derived basically from a statistical model approach
\cite{Mazurek} and are orders of magnitude too small \cite{Brown89}.
What are the consequences and will electron capture and $\beta$ decay
rate also balance in a consistent model?  At the beginning of the
collapse, electron capture is significantly faster than $\beta$ decay
(see Fig. 3). The shell model rates make $^{56}$Ni the most important
contributor, but it cannot quite compensate for the reduction of the
$^{55}$Co rate. Thus, at $Y_e=0.485$ the total electron capture rate
${\dot Y}_e^{ec}$ drops slightly. This reduction is more severe for
smaller $Y_e$ values, until at $Y_e=0.46$ electron capture and $\beta$
decay balance. The consequence is that, due to the slower electron
capture, the star radiates less energy away in form of neutrinos until
$Y_e=0.46$ is reached.  Thus one expects that in the early stage the
stellar trajectory is, for a given density, at a higher temperature.
This, of course, increases both the $\beta$ decay and electron capture
rates.  Importantly both rates have roughly the same temperature
dependence in the relevant temperature range: typically electron
capture rates are enhanced by an order of magnitude if the temperature
raises from $T_9=4$ to $T_9=6$. But this increase is the same order of
magnitude by which the $\beta$ decay rates grow in the same
temperature interval. Consequently the two rates will also be balanced
at around $Y_e \approx 0.46$ if a consistent stellar trajectory is
used.

As stated above, the dominance of $\beta$ decay over electron capture
during a certain stage of the core collapse of a massive star has been
suggested or noted before
\cite{Brown89,Aufderheide,Aufderheide94,Aufderheide96}.  However,
previous argumentation has been based on rates for these two processes
which had been empirically and intuitively parametrized, rather than
derived from a reliable many-body model. Moreover, it was shown in
recent years that the assumed parametrizations, mainly with respect to
the energy of the Gamow-Teller centroid, were systematically
incorrect. Shell model calculations are now at hand which allow, for
the first time, the reliable calculation of these rates under stellar
conditions.  Given the fact that the large-scale shell model studies
reproduce all important ingredients (spectra, half-lives, GT strength
distributions) very well, the shell model rates are rather reliable.
We stress an important point, that the shell model $\beta$ decay rates
are larger than the electron capture rates for $Y_e \approx
0.42-0.455$. This might have important consequences for the core
collapse. A first investigation into these consequences has been
performed by Aufderheide et al.  \cite{Aufderheide94}, however, using
the FFN values for both rates.  They find that the competition of
$\beta$ decay and electron capture leads to cooler cores and larger
$Y_e$ values at the formation of the homologuous core.  These results
are important motivation enough to derive a complete set of shell
model rates and then use them in core collapse calculations.

\acknowledgements

This work was supported in part by the Danish Research Council and
through grant DE-FG02-96ER40963 from the U.S. Department of Energy.
Oak Ridge National Laboratory is managed by Lockheed Martin Energy
Research Corp. for the U.S. Department of Energy under contract number
DE-AC05-96OR22464..  Grants of computational resources were provided
by the Center for Advanced Computational Research at Caltech.  KL and
GMP thank Michael Strayer and the Oak Ridge Physics Division for their
kind hospitalty during which parts of the manuscript have been
written.

\end{multicols}

\newpage

\begin{table}
  \begin{center}
    \caption{Electron capture  rates for selected even-even, odd-A and
    odd-odd nuclei. The chosen stellar conditions reflect those at
    which the nuclei are considered to be most important due to the
    ranking given by Aufderheide et al. \protect\cite{Aufderheide}.
    The shell model rates (labelled SM) are compared to those
    recommended by FFN \protect\cite{FFN} and Ref.
    \protect\cite{Aufderheide}. The last column, named importance
    ratio, gives the percentage of the total change in ${\dot Y}_e$
    (for the definition see text) assigned to the respective nucleus
    by Aufderheide et al. at the respective stellar conditions. All
    rates are in s$^{-1}$.  Exponents are given in parentheses.}
\begin{tabular}{ccccccc}
nucleus & $\rho_7$ & $T_9$ & SM & FFN & Ref. \protect\cite{Aufderheide}
& importance ratio \\ \hline
$^{56}$Ni & 4.32 & 3.26 & 1.3 (-2) & 7.4 (-3) & 8.6 (-3) & 0.254 \\
$^{54}$Fe & 5.86 & 3.40 & 4.2 (-5) & 2.9 (-4) & 3.1 (-4) & 0.126 \\ 
$^{58}$Ni & 5.86 & 3.40 & 8.1 (-5) & 3.7 (-4) & 6.3 (-4) & 0.065 \\
$^{56}$Fe & 10.7 & 3.65 & 2.1 (-6) & 1.0 (-5) & 4.7 (-7) & 0.005 \\
\hline
$^{55}$Co & 4.32 & 3.26 & 1.6 (-3) & 8.4 (-2) & 5.1 (-2) & 0.501 \\
$^{57}$Co & 5.86 & 3.40 & 1.3 (-4) & 1.9 (-3) & 3.4 (-3) & 0.246 \\
$^{55}$Fe & 5.86 & 3.40 & 1.9 (-4) & 5.8 (-3) & 3.8 (-3) & 0.126 \\
$^{59}$Ni & 5.86 & 3.40 & 4.7 (-4) & 4.4 (-3) & 4.4 (-3) & 0.041 \\
$^{59}$Co & 10.7 & 3.65 & 7.8 (-6) & 2.1 (-4) & 2.1 (-4) & 0.151 \\
$^{53}$Mn & 10.7 & 3.65 & 3.3 (-4) & 3.8 (-3) & 5.6 (-3) & 0.097 \\
\hline
$^{56}$Co & 5.86  & 3.40 & 1.7 (-3) & 6.9 (-2) & 5.1 (-2) & 0.074 \\
$^{54}$Mn & 10.7  & 3.65 & 3.1 (-4) & 4.5 (-3) & 1.1 (-2) & 0.188 \\
$^{58}$Co & 10.7  & 3.65 & 3.5 (-4) & 9.1 (-3) & 2.1 (-2) & 0.057 \\
$^{56}$Mn & 33.0  & 4.24 & 1.0 (-4) & 4.1 (-4) & 2.0 (-3) & 0.058 \\
$^{60}$Co & 33.0  & 4.24 & 1.7 (-4) & 1.1 (-1) & 6.1 (-2) & 0.513 \\
\end{tabular}
\end{center}
\end{table}

\begin{table}
\begin{center}
\caption{$\beta$ decay rates for selected even-even, odd-A and odd-odd
  nuclei. The chosen stellar conditions reflect those at which the
  nuclei are considered to be most important due to the ranking given
  by Aufderheide et al. \protect\cite{Aufderheide}. The shell model
  rates (labelled SM) are compared to those recommended by FFN
  \protect\cite{FFN} and in Ref. \protect\cite{Aufderheide}. The last
  column, labelled importance ratio, gives the percentage of the total
  change in ${\dot Y}_e$ (for the definition see text) assigned to the
  respective nucleus by Aufderheide et al. at the respective stellar
  conditions. All rates are in s$^{-1}$.  Exponents are given in
  parentheses. FFN did not give rates for nuclei with $A>60$.}
\begin{tabular}{ccccccc}
  nucleus & $\rho_7$ & $T_9$ & SM & FFN & Ref. \protect\cite{Aufderheide}
  & importance ratio \\ \hline
  $^{56}$Fe & 5.86 & 3.40 & 3.9 (-11) & 2.3 (-10) & 5.9 (-11) & 0.006 \\
  $^{54}$Cr & 5.86 & 3.40 & 2.2 (-7)  & 2.2 (-5) & 1.5 (-7) & 0.032 \\ 
  $^{58}$Fe & 10.7  & 3.65 & 5.2 (-8) & 2.6 (-6) & 1.5 (-7) & 0.004 \\
  $^{60}$Fe & 33.0  & 4.24 & 1.7 (-4) & 4.6 (-3) & 1.0 (-3) & 0.112 \\
  $^{52}$Ti & 33.0  & 4.24 & 1.3 (-3) & 1.1 (-2) & 1.2 (-4) & 0.001 \\
  \hline
  $^{59}$Fe & 33.0  & 4.24 & 6.0 (-5) & 6.3 (-3) & 5.3 (-3) & 0.245 \\
  $^{61}$Fe & 33.0  & 4.24 & 1.7 (-3) &          & 6.4 (-2) & 0.126 \\
  $^{61}$Co & 33.0  & 4.24 & 1.6 (-4) &          & 9.3 (-4) & 0.029 \\
  $^{63}$Co & 33.0  & 4.24 & 1.6 (-2) &          & 1.4 (-2) & 0.057 \\
  $^{59}$Mn & 220   & 5.39 & 2.2 (-2) & 7.2 (-1) & 1.4 (-1) & 0.095 \\
  \hline
  $^{58}$Co & 4.32  & 3.26 & 2.7 (-6) & 1.2 (-6) & 3.8 (-6) & 0.096 \\
  $^{54}$Mn & 5.86  & 3.40 & 2.7 (-6) & 1.6 (-6) & 7.5 (-6) & 0.320 \\
  $^{56}$Mn & 10.7  & 3.65 & 3.4 (-3) & 3.0 (-3) & 9.1 (-3) & 0.235 \\
  $^{60}$Co & 10.7  & 3.65 & 6.6 (-4) & 1.4 (-3) & 3.4 (-3) & 0.116 \\
  $^{50}$Sc & 33.0  & 4.24 & 1.2 (-2) & 2.8 (-2) & 1.8 (-1) & 0.025 \\
\end{tabular}
\end{center}
\end{table}

\begin{figure}
  \begin{center}
    \leavevmode
    \epsfxsize=0.8\columnwidth
    \epsffile{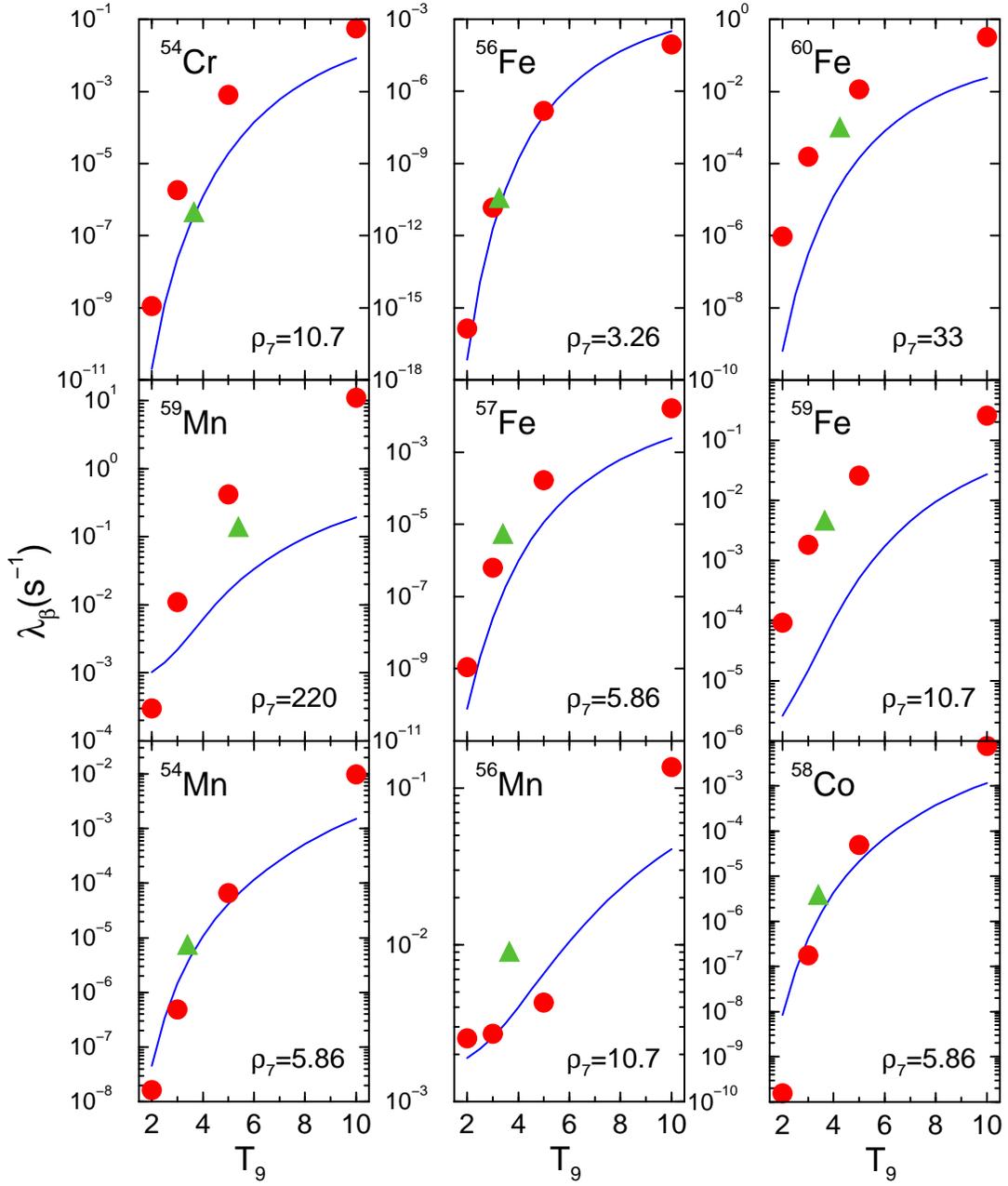}
    \caption{$\beta$ decay rates for several nuclei as a function of
      temperature and at selected densities at which these nuclei are
      important for electron capture in the presupernova core collapse
      as suggested by Ref. \protect\cite{Aufderheide}. The top
      (middle, bottom) row contains even-even (odd-A, odd-odd) nuclei.
      The solid line shows the present shell model results, the dots
      give the FFN rates \protect\cite{FFN}, while the triangles are
      rates taken from Tables 15-17 in \protect\cite{Aufderheide}.  }
    \label{fig1}
  \end{center}
\end{figure}

\begin{figure}
  \begin{center}
    \leavevmode
    \epsfxsize=0.8\columnwidth
    \epsffile{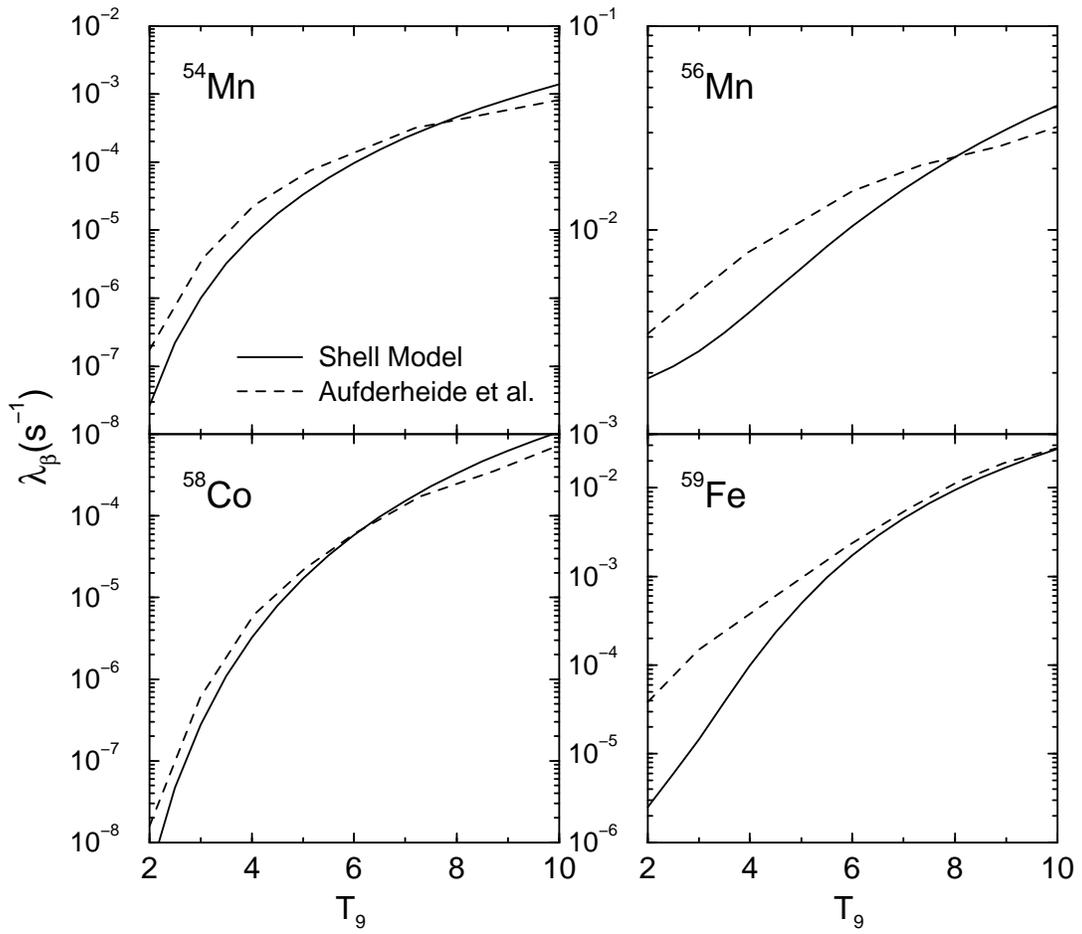}
    \caption{Comparison of the present shell model rates for
      $^{54,56}$Mn, $^{58}$Co and $^{59}$Fe (solid line) with those
      derived in Ref.  \protect\cite{Aufderheide96} (dashed line).  }
    \label{fig2}
  \end{center}
\end{figure}

\begin{figure}
  \begin{center}
    \leavevmode
    \epsfxsize=0.8\columnwidth
    \epsffile{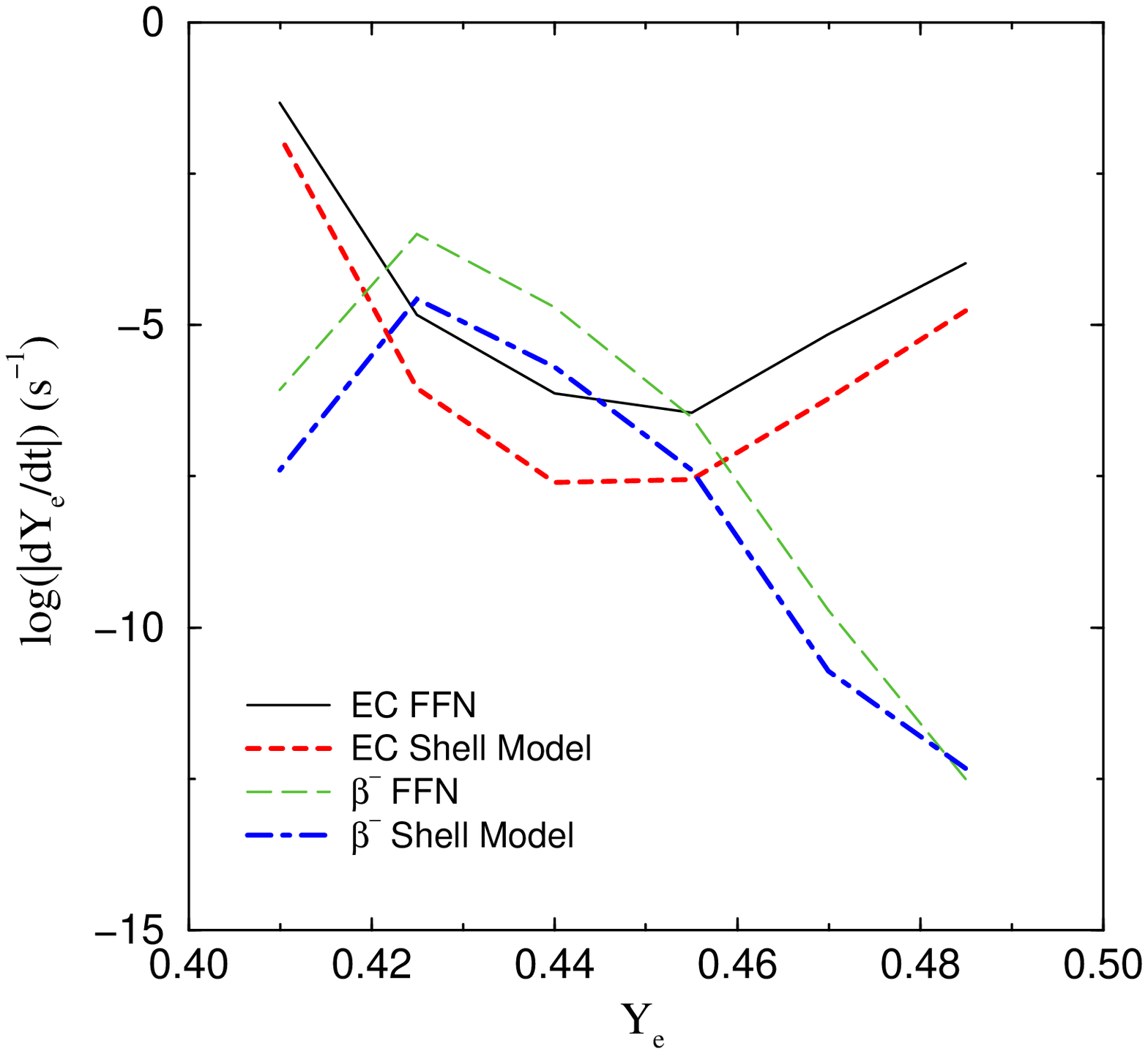}
    \caption{Change in the total electron capture and $\beta$ decay
      rates, ${\dot Y}_e^{ec}$ and ${\dot Y}_e^{\beta}$, respectively.
      The shell model results are compared with the FFN results along
      the same stellar trajectory as in Fig. 14 of Ref.
      \protect\cite{Aufderheide}.}
    \label{fig3}
  \end{center}
\end{figure}

\begin{references}
  
\bibitem{Bethe78} H.A. Bethe, G.E. Brown, J. Applegate and J.M.
  Lattimer, Nucl. Phys. {\bf A324} (1979) 487
  
\bibitem{Bethe90} H.A. Bethe, Rev. Mod. Phys. {\bf 62} (1990) 801
  
\bibitem{Thielemann} F.-K. Thielemann, private communication
  
\bibitem{Brown89} G.E. Brown, proceedings of the International Nuclear
  Physics Conference 1989, eds. M.S. Hussein et al. (World Scientific,
  Singapore, 1990) p. 3
  
\bibitem{Aufderheide} M. B. Aufderheide, I. Fushiki, S. E. Woosley,
  and D. H.  Hartmann, Astrophys. J. Suppl. 91 (1994) 389

\bibitem{FFN} G.M. Fuller, W.A. Fowler and M.J. Newman, ApJS {\bf 42}
  (1980) 447; {\bf 48} (1982) 279; ApJ {\bf 252} (1982) 715; {\bf 293}
  (1985) 1
  
\bibitem{gtdata1} A.L. Williams {\it et al.}, Phys. Rev. {\bf C51}
  (1995) 1144
  
\bibitem{gtdata2} W.P. Alford {\it et al.}, Nucl. Phys. {\bf A514}
  (1990) 49 \bibitem{gtdata3} M.C. Vetterli {\it et al.}, Phys. Rev.
  {\bf C40} (1989) 559
  
\bibitem{gtdata4} S. El-Kateb {\it et al.}, Phys. Rev. {\bf C49}
  (1994) 3129
  
\bibitem{gtdata5} T. R\"onnquist {\it et al.}, Nucl. Phys. {\bf A563}
  (1993) 225
  
\bibitem{Aufderheide96} M.B. Aufderheide, S.D. Bloom, G.J. Mathews and
  D.A. Resler, Phys. Rev. {\bf C53} (1996) 3139
  
\bibitem{Dean98} D.J. Dean, K. Langanke, L. Chatterjee, P.B. Radha and
  M.R. Strayer, Phys. Rev. {\bf C 58} (1998) 536
  
\bibitem{Langanke98a} K. Langanke and G. Mart\'{\i}nez Pinedo, Phys.
  Lett.  B436 (1998) 19
  
\bibitem{Langanke98b} K. Langanke and G. Mart\'{\i}nez-Pinedo,
  submitted to Phys. Lett. B
  
\bibitem{Martinez99} G. Mart\'{\i}nez-Pinedo, E. Caurier, K. Langanke
  and F.  Nowacki, in preparation
  
\bibitem{Caurier98} E. Caurier, G. Mart\'{\i}nez-Pinedo, F. Nowacki
  and A.  Poves, LANL archive nucl-th/9809068, submitted to Phys. Rev.
  C.
  
\bibitem{Johnson} C.W. Johnson, S.E. Koonin, G.H. Lang and W.E.
  Ormand, Phys. Rev. Lett. {\bf 69} (1992) 3157
  
\bibitem{Koonin} S.E. Koonin, D.J. Dean and K. Langanke, Phys. Rep.
  {\bf 278} (1996) 1
  
\bibitem{Ni56} D. Rudolph et al., submitted to Phys. Rev. Lett.
  
\bibitem{Wildenthal} B.A Brown and B.H. Wildenthal, {Ann. Rev. Nucl.
    Part. Sci.} {\bf 38}, 29 (1988).
  
\bibitem{Langanke95} K. Langanke, D.~J. Dean, P.~B. Radha, Y.
  Alhassid, and S.~E. Koonin, {Phys. Rev. C} {\bf 52}, 718 (1995).
  
\bibitem{Martinez96} G. Mart\'{\i}nez-Pinedo, A. Poves, E. Caurier,
  and A.~P. Zuker, {Phys. Rev. C} {\bf 53}, R2602 (1996).
  
\bibitem{Aufderheide91} M.B. Aufderheide, Nucl. Phys. {\bf A526}
  (1991) 161
  
\bibitem{Aufderheide90} M.B. Aufderheide, G.E. Brown, T.T.S. Kuo, D.B.
  Stout and P. Vogel, ApJ. {\bf 362} (1990) 241

\bibitem{Kepler} T.A. Weaver, G.B. Zimmerman and S.E. Woosley, ApJ.
  {\bf 225} (1978) 1021
  
\bibitem{Aufderheide94} M.B. Aufderheide, I. Fushiki, G.M. Fuller and
  T.A. Weaver, ApJ. {\bf 424} (1994) 257

\bibitem{Mazurek} T.J. Mazurek, J.W. Truran and A.G.W. Cameron, ApSS
  {\bf 27} (1974) 261
  
\end{references}
\end{document}